\documentclass[12pt]{article}
\usepackage[dvips]{color}
\usepackage{epsfig}
\usepackage{amsmath}
\usepackage{graphicx}
\def\Box{\hbox{$\rlap{$\sqcup$}\sqcap$}}
\begin{document}
\setcounter{page}{1}

\pagestyle{plain} \vspace{1cm}

\begin{center}
\Large{\bf Early and Late-time Cosmic Acceleration in Non-minimal
Yang-Mills-$f(G)$ Gravity }\\
\small \vspace{1cm}{\bf A. Banijamali $^{a,}$
\footnote{a.banijamali@nit.ac.ir}} and {\bf B. Fazlpour $^{b,}$
\footnote{b.fazlpour@umz.ac.ir}}  \\
\vspace{0.5cm} $^{a}$ {\it Department of Basic Sciences, Babol
University of Technology, Babol, Iran\\}\vspace{0.5cm} $^{b}$ {\it
Department of Physics, Ayatollah Amoli Branch, Islamic Azad
University, P. O. Box 678, Amol, Iran\\}
\end{center}
\vspace{1.5cm}
\begin{abstract}
In this paper we show that power-law inflation can be realized in
non-minimal gravitational coupling of Yang-Mills field with a
general function of the Gauss-Bonnet invariant in the framework of
Einstein gravity. Such a non-minimal coupling may appear due to
quantum corrections. We also discuss the non-minimal
Yang-Mills-$f(G)$ gravity in the framework of modified Gauss-Bonnet
action which is widely studied recently. It is shown that both
inflation and late-time cosmic acceleration are possible in such a
theory.

 \noindent
\hspace{0.35cm}

\end{abstract}
\newpage
\section{Introduction}
Inflation in the early time and  acceleration in the current
expansion of the universe are two periods of accelerated expansion
in our universe which are confirmed by cosmological observations
[1-4]. There are two ways to explain the current accelerated
expansion of the universe. The first one is introducing some unknown
matter, which is called dark energy [5] in the framework of general
relativity
and the second one is modified gravity.\\
The modified gravity in the simplest type, uses an arbitrary
function $f$ of the Ricci scalar instead of $R$ in Einstein-Hilbert
action which is  known as $f(R)$ gravity  (see [6-8] for reviews).
There are also other modified gravity models which are the
generalizations of $f(R)$ gravity and among them, the modified
Gauss-Bonnet (GB) gravity i.e. $f(G)$ gravity, is more interesting
[9,10]. In order to play some role in the Friedmann equation, it is
required that the GB combination $G$, a topological invariant in
four dimensions,  couples to a scalar field $\phi$ or the Lagrangian
density be a function of $G$ i.e $f(G)$. The GB coupling with a
scalar field appears in the low energy effective action of
string/M-theory [11] and the cosmological solution in such a theory
have been studied in great details [12]. It was shown that, if the
GB term is responsible for DE,
this model does not satisfy local gravity constraints easily [13].\\
Furthermore, the modified $f(G)$ gravity has the possibility to
describe the inflationary era, a transition from a deceleration
phase to an acceleration phase, crossing the phantom divide line and
passing the solar system tests for a reasonable defined function $f$
[14-16]. The $f(G)$ models might be less constrained by local
gravity constraints compared to the $f(R)$ models [17]. Hence
modified $f(G)$ gravity represents a quite interesting gravitational
alternative for dark energy (for a recent review see [8]).\\
The non-minimal coupling of the Ricci scalar and matter Lagrangian, can
be seen as a source of inflation and late-time accelerated expansion
of the universe [18-20]. Such a non-minimal coupling has been
studied widely in the literature. For example, the non-minimal
coupling between $f(R)/f(G)$ gravity and the kinetic part of
Lagrangian of a massless scalar field has been investigated in Ref.
[21]. Non-minimal coupling of a viable model of $f(R)$ gravity and
electromagnetic Lagrangian has been considered in Ref. [22] and it
has been shown that inflation and current cosmic acceleration can be
explained in such a model. The coupling between scalar curvature and
Lagrangian of the electromagnetic field arises in curved spacetime
due to one-loop vacuum polarization effects in Quantum Electro
Dynamics (QED) [23] and breaks the conformal invariance of the
electromagnetic field, so that electromagnetic quantum fluctuations
can be generated at the inflationary stage and they act
as a source for inflation.\\
It has been shown that both inflation and late-time accelerated
expansion of the universe can be realized in a modified non-minimal
Yang-Mills-$f(R)$ gravity [24]. Also, this result can be realized in
a non-minimal vector-$f(R)$ gravity in the framework of modified
gravity [24]. Moreover, the conditions for the non-minimal
gravitational coupling of the electromagnetic field in order to
remove the finite-time singularities  have been investigated in Ref.
[25]. Ref. [26] has  considered $f(R)$ gravity coupling to
non-linear electrodynamics. The criteria for the validity of a
non-minimal coupling between scalar curvature and
matter Lagrangian have been studied in [27-30].\\
Furthermore, realizing power-law inflation in non-minimal
gravitational coupling of electromagnetic field with a general
function of Gauss-Bonnet invariant has been done in [31]. Also, it
has been demonstrated that both inflation and late-time acceleration
of the universe can be realized in a modified Maxwell-$f(G)$ gravity
proposed in Ref. [32] in the framework of
modified Gauss-Bonnet gravity.\\
In this paper we study early and late-time cosmic acceleration in
non-minimal Yang-Mills-$f(G)$ gravity in which Yang Mills (YM) field
couples to a function of Gauss-Bonnet invariant. Non-minimal
coupling appears in some string compactification where extra
curvature terms exist in front of YM Lagrangian. Also, since the
energy scale of the YM theory is higher than the electroweak scale,
the existence of YM field with a non- minimal gravitational coupling
might have influence on models of grand unified theories (GUT) [24].
In the past studies, the effective YM condensation as a candidate
for dark energy has been proposed in [33,34] and the possibility for
cosmic acceleration  driven by a field with an anisotropic equation
of state has been studied in [35]. We show that power-law inflation
can be realized in non-minimal gravitational coupling of the YM
field in the Einstein frame. Additionally, we demonstrate that both
inflation and late-time acceleration of the universe can be realized
in YM-$f(G)$ model in the framework of modified Gauss-Bonnet gravity
in which the function for $f(G)$ is consistent with the solar system tests.\\
An outline of this paper is as follows. In section 2 we examine
power-law inflation in a model of non-minimally coupled YM field
with $f(G)$ gravity in the general relativity (GR) framework. In
section 3 we show that both inflation and late-time cosmic
acceleration can be realized in such a model but in the framework of
modified Gauss-Bonnet gravity proposed in [32]. Section 4 is devoted
to our conclusion.

\section{Power-law inflation in non-minimal Yang- Mills gravity}

Our starting action is as follows:

\begin{eqnarray}
S=\int d^{4}x
\sqrt{-g}\Bigg[\frac{1}{2\kappa^{2}}R-\frac{1}{4}\big(F_{\mu\nu}^{a}F^{a\mu\nu}+f(G)F_{\mu\nu}^{a}F^{a\mu\nu}\big)\Big[1+b\tilde{g}^{2}\ln
\Big|\frac{-(1/2)F_{\mu\nu}^{a}F^{a\mu\nu}}{\mu^{4}}\Big|\Big]\Bigg],
\end{eqnarray}
where $g$ is the determinant of metric tensor $g_{\mu\nu}$, $R$ is
the Ricci scalar and $f(G)$ is a general function of Gauss- Bonnet
invariant, $G$, which is non-minimally coupled with YM Lagrangian.
The effective YM Lagrangian up to one-loop order is as follows
[36, 37]:
\begin{eqnarray}
-\frac{1}{4}\big(F_{\mu\nu}^{a}F^{a\mu\nu}+f(G)F_{\mu\nu}^{a}F^{a\mu\nu}\big)\Big[1+b\tilde{g}^{2}\ln
\Big|\frac{-(1/2)F_{\mu\nu}^{a}F^{a\mu\nu}}{\mu^{4}}\Big|\Big].
\end{eqnarray}
The field strength tensor of YM Lagrangian is
$F_{\mu\nu}^{a}=\partial_{\mu}A_{\nu}^{a}-\partial_{\nu}A_{\mu}^{a}+f^{abc}A_{\mu}^{b}A_{\nu}^{c}$
, where $A_{\mu}^{a}$ is the $SU(N)$ YM field and $f^{abc}$ is the
structure constants and completely antisymmetric [38]. Moreover, $b$
is a constant and $\tilde{g}$ is also constant which is a function
of field strength and $\mu$ is the mass scale of the renormalization
point [37]. We note that the action (1) without the third term,
corresponds to usual Einstein-YM theory. Varying the action (1) with
respect to $A_{\mu}^{a}$ leads to the following equations of motion
for $SU(N)$ YM field,
\begin{equation}
\frac{1}{\sqrt{-g}}\partial_{\mu}\Big[\sqrt{-g}\big(1+f(G)\big)\varepsilon
F^{a\mu\nu}\Big]-\big(1+f(G)\big)\varepsilon
f^{abc}A_{\mu}^{b}F^{c\mu\nu}=0,
\end{equation}
where $\varepsilon$ is the effective dielectric constant which
depends on field strength [37]. Now variation of (1) with respect to
the metric $g_{\mu\nu}$ leads to
\begin{eqnarray}
0=\frac{1}{\sqrt{-g}}\frac{\delta S}{\delta
g^{\mu\nu}}=\frac{1}{2\kappa^{2}}(\frac{1}{2}g_{\mu\nu}R-R_{\mu\nu})+T_{\mu\nu}^{eff}.
\end{eqnarray}
Here the effective energy momentum tensor $T_{\mu\nu}^{eff}$ is
defined by
\begin{eqnarray}
T_{\mu\nu}^{eff}&=&\frac{1}{2}(1+f(G))\Big(\varepsilon
g^{\gamma\delta}F_{\mu\delta}^{a}F_{\nu\gamma}^{a}-\frac{1}{4}
g_{\mu\nu}F_{\gamma\delta}^{a}F^{a\gamma\delta}\mathcal{B}\Big)
+\frac{1}{2}\Big\{f'(G)F_{\gamma\delta}^{a}F^{a\gamma\delta}\mathcal{B}R\,R_{\mu\nu}
\nonumber\\
&-&2f'(G)F_{\gamma\delta}^{a}F^{a\gamma\delta}\mathcal{B}R_{\mu}\,^{\rho}R_{\nu\rho}+
f'(G)F_{\gamma\delta}^{a}F^{a\gamma\delta}\mathcal{B}R_{\mu\rho\sigma\lambda}R_{\nu}\,^{\rho\sigma\lambda}
+2f'(G)F_{\gamma\delta}^{a}F^{a\gamma\delta}\mathcal{B}R_{\mu\rho\sigma\nu}R^{\rho\sigma}
\nonumber\\ &+& [(g_{\mu\nu}\Box-\nabla_{\mu}\nabla_{\nu})
f'(G)F_{\gamma\delta}^{a}F^{a\gamma\delta}\mathcal{B}]R
+2[\nabla^{\rho}\nabla_{\mu}
(f'(G)F_{\gamma\delta}^{a}F^{a\gamma\delta}\mathcal{B})]R_{\nu\rho}\nonumber\\
&+&2[\nabla^{\rho}\nabla_{\nu}
(f'(G)F_{\gamma\delta}^{a}F^{a\gamma\delta}\mathcal{B})]R_{\mu\rho}
+2[\Box
(f'(G)F_{\gamma\delta}^{a}F^{a\gamma\delta}\mathcal{B})]R_{\mu\nu}\nonumber\\
&-&2[g_{\mu\nu}\nabla^{\lambda}
\nabla^{\rho}(f'(G)F_{\gamma\delta}^{a}F^{a\gamma\delta}\mathcal{B})]
R_{\lambda\rho}+2[\nabla^{\rho}
\nabla^{\lambda}(f'(G)F_{\gamma\delta}^{a}F^{a\gamma\delta}\mathcal{B})]R_{\mu\rho\nu\lambda}\Big\},
\end{eqnarray}
where
\begin{equation}
\mathcal{B}=\Big[1+b\tilde{g}^{2}\ln
\Big|\frac{-(1/2)F_{\gamma\delta}^{a}F^{a\gamma\delta}}{\mu^{4}}\Big|\Big],
\end{equation}
and
\begin{equation}
\varepsilon=1+b\tilde{g}^{2}\ln
\Big|e\Big[\frac{-(1/2)F_{\gamma\delta}^{a}F^{a\gamma\delta}}{\mu^{4}}\Big]\Big|.
\end{equation}
In equation (5), $f'(G)=\frac{df(G)}{dG}$,
$\Box=g^{\mu\nu}\nabla_{\mu}\nabla_{\nu}$ is the d'Alembertian
operator and $e\approx 2.72$ is the Napierian number.\\
In a flat Friedmann-Robertson-Walker (FRW) spacetime with the
metric
\begin{eqnarray}
ds^{2}=-dt^{2}+a^{2}(t)(dr^{2}+r^{2}d\Omega^{2}),
\end{eqnarray}
the  components of the Ricci tensor $R_{\mu\nu}$ and the Ricci scalar $R$
are given by
\begin{equation}
R_{00}=-3\big(\dot{H}+H^{2}\big),\,\,R_{ij}=a^{2}(t)\big(\dot{H}+3H^{2}\big)\delta_{ij},\,\,
R=6\big(\dot{H}+2H^{2}\big),
\end{equation}
where $H=\frac{\dot{a}(t)}{a(t)}$ is the Hubble parameter and $a(t)$
is the scale factor. Also the Gauss- Bonnet invariant in this background
is
\begin{equation}
 G=24\big(\dot{H}H^{2}+H^{4}\big).
\end{equation}
The $(0,0)$ component and sum of $(i,i)$ components of equation (4)
in FRW spacetime have the following forms respectively
$$H^{2}=\frac{\kappa^{2}}{3}\Bigg[\frac{1}{2}\big(1+f(G)\big)\big(b\tilde{g}^{2}\mathcal{X}+\varepsilon\mathcal{Y
}\big)+6\Big(f'(G)\big(\dot{H}H^{2}+H^{4}\big)$$
\begin{equation}
-24H^{3}\big(\ddot{H}H^{2}+2H\dot{H}^{2}+4\dot{H}
H^{3}\big)f''(G)\Big)F_{\gamma\delta}^{a}F^{a\gamma\delta}\mathcal{B}-6H^{3}f'(G)\frac{\partial}{\partial
t} \big(F_{\gamma\delta}^{a}F^{a\gamma\delta}\mathcal{B}\big)\Bigg],
\end{equation}
and
\begin{eqnarray}
2\dot{H}+3H^{2}&=&\kappa^{2}\Bigg[\frac{1}{2}\big(1+f(G)\big)\mathcal{X}
\big(-\frac{1}{3}\varepsilon+b\tilde{g}^{2}\big)+
\Bigg(6f'(G)\big(\dot{H}H^{2}+H^{4}\big)\nonumber\\
 &-&24\Big[2f''(G)\big(8\ddot{H}\dot{H}H^{3}+6\dot{H}^{3}H^{2}
 +24 \dot{H}^{2}H^{4}+6\ddot{H}H^{5}+8\dot{H}^{2}H^{6}+\dddot{H}H^{4}\big)\nonumber\\
 &+&f'''(G)
 \big(\ddot{H}H^{2}+2H\dot{H}^{2}+4\dot{H}H^{3}\big)^{2}\Big]
 \Bigg)F_{\gamma\delta}^{a}F^{a\gamma\delta}\mathcal{B}-4
 \Big[f'(G)\big(H\dot{H}+H^{3}\big)\nonumber\\&+&24f''(G)\big(\ddot{H}H^{4}+2\dot{H}^{2}H^{3}+4\dot{H}H^{5}\big)
 \Big]\frac{\partial}{\partial t}\big(F_{\gamma\delta}^{a}F^{a\gamma\delta}\mathcal{B}\big)-2f'(G)H^{2}\frac{\partial^{2}}
 {\partial
 t^{2}}\big(F_{\gamma\delta}^{a}F^{a\gamma\delta}\mathcal{B}\big)\Bigg],\nonumber\\
\end{eqnarray}
where we have neglected the second order spatial derivative of the
quadratic quantity
  $F_{\gamma\delta}^{a}F^{a\gamma\delta}$.\\
In addition, $\mathcal{X}$ and $\mathcal{Y}$ in above equations are
given by [24],
\begin{equation}
\mathcal{X}=|E_{i}^{a}(t)|^{2}-|B_{i}^{a}(t)|^{2},
\end{equation}
\begin{equation}
\mathcal{Y}=|E_{i}^{a}(t)|^{2}+|B_{i}^{a}(t)|^{2},
\end{equation}
and
\begin{equation}
F_{\gamma\delta}^{a}F^{a\gamma\delta}\mathcal{B}=-2\mathcal{X}\big(\varepsilon-b\tilde{g}^{2}\big).
\end{equation}
 Here $E_{i}^{a}(t)$ and $B_{i}^{a}(t)$ are proper electric and magnetic fields in the $SU(N)$ YM theory
 respectively.\\
 As a simple case our interest is the generation of large scale (YM) magnetic fields instead of (YM)
electric fields, so we neglect terms in (YM) electric fields from
this point and hence we have, $\mathcal{X}=-|B_{i}^{a}(t)|^{2}$ and
$\mathcal{Y}=|B_{i}^{a}(t)|^{2}$. Furthermore, for more simplicity,
we assume that only one component of (YM) magnetic fields is
non-zero and the other two components are zero i.e.
$B_{1}^{a}=B_{2}^{a}=0$, $B_{3}^{a}\neq0$. We note that by
considering these conditions, the off-diagonal components of the
last term on the right hand side of $T_{\mu\nu}^{eff}$ are zero and
so, all
off-diagonal components of $T_{\mu\nu}^{eff}$ are zero.\\
 The amplitude of (YM) magnetic fields on a comoving scale $L=\frac{2\pi}{k}$ with
 the comoving wave
 number $k$ in the position space is given by
\begin{eqnarray}
|B_{i}^{a}(t)|^{2}=\frac{|B_{0}^{a}|^{2}}{a^{4}},
\end{eqnarray}
where $|B_{0}^{a}|$ is a constant. Now substituting (16) in (11) and
(12) and using equations (13)-(15) leads to
$$H^{2}=\kappa^{2}\Bigg[\frac{1}{6}\big(1+f(G)\big)\big(\varepsilon-b\tilde{g}^{2}\big)+4\Big(f'(G)\big(\dot{H}H^{2}+H^{4}\big)$$
\begin{equation}
-24H^{3}\big(\ddot{H}H^{2}+2H\dot{H}^{2}+4\dot{H}
H^{3}\big)f''(G)\Big)\big(\varepsilon-b\tilde{g}^{2}\big)+16H^{4}f'(G)\varepsilon\Bigg]\frac{|B_{0}^{a}|^{2}}{a^{4}},
\end{equation}
and
\begin{eqnarray}
2\dot{H}+3H^{2}&=&\kappa^{2}\Bigg[\frac{1}{6}\big(1+f(G)\big)
\big(\varepsilon-3b\tilde{g}^{2}\big)+2
\Bigg(6f'(G)\big(\dot{H}H^{2}+H^{4}\big)\nonumber\\
 &-&24\Big[2f''(G)\big(8\ddot{H}\dot{H}H^{3}+6\dot{H}^{3}H^{2}
 +24 \dot{H}^{2}H^{4}+6\ddot{H}H^{5}+8\dot{H}^{2}H^{6}+\dddot{H}H^{4}\big)\nonumber\\
 &+&f'''(G)
 \big(\ddot{H}H^{2}+2H\dot{H}^{2}+4\dot{H}H^{3}\big)^{2}\Big]
 \Bigg)\big(\varepsilon-b\tilde{g}^{2}\big)+32
 \Big[f'(G)\big(\frac{3}{2}H^{2}\dot{H}-H^{4}\big)\nonumber\\&+&24f''(G)H\big(\ddot{H}H^{4}+2\dot{H}^{2}H^{3}+4\dot{H}H^{5}\big)
 \Big]\varepsilon+64f'(G)H^{4}b\tilde{g}^{2}\Bigg]\frac{|B_{0}^{a}|^{2}}{a^{4}},\nonumber\\
\end{eqnarray}
respectively. Then from equations (17) and (18) one can obtain
$$\dot{H}+\frac{\varepsilon}{\varepsilon-b\tilde{g}^{2}}H^{2}=\kappa^{2}\Bigg[4f'(G)\Big\{7\dot{H}H^{2}\varepsilon-\Big[
\Big(\frac{5\varepsilon-9b\tilde{g}^{2}}{\varepsilon-b\tilde{g}^{2}}\Big)\varepsilon-8b\tilde{g}^{2}\Big]H^{4}\Big\}+48f''(G)
\Big\{3\ddot{H}H^{5}(\varepsilon+b\tilde{g}^{2})$$
$$-6H^{4}\dot{H}^{2}(\varepsilon-3b\tilde{g}^{2})+4\dot{H}H^{6}(7
\varepsilon-b\tilde{g}^{2})-\big(8\ddot{H}\dot{H}H^{3}+6\dot{H}^{3}H^{2}
+\dddot{H}H^{4}\big)(\varepsilon-b\tilde{g}^{2})\Big\}+24f'''(G)
 \big(\ddot{H}H^{2}$$
\begin{equation}
+2H\dot{H}^{2}+4\dot{H}H^{3}\big)^{2}(\varepsilon-b\tilde{g}^{2})\Bigg]\frac{|B_{0}^{a}|^{2}}{a^{4}}.\hspace{10cm}
\end{equation}
From this point, we are going to consider $\varepsilon$ as a
constant. The reason for doing so, is that the dependence of
$\varepsilon$ on the field strength and therefore on time, is
logarithmic as one can see from equation (7), while this is not the
case for the other quantities in above equation.\\
Now we examine the following function for $f(G)$, which has been
proposed in Ref. [21]:
\begin{equation}
f(G)=\frac{G^{n}}{c_{1}G^{n}+c_{2}},
\end{equation}
where  $c_{1}$ and $c_{2}$ are constants and $n$ is a positive
integer. It is known [32] that such a model naturally leads to
unification of the inflation with late-time acceleration being
consistent with local tests and cosmological bounds. We can see in
the late-time universe, the ordinary YM theory can be naturally
recovered because the value of Gauss- Bonnet invariant in this time
goes to zero.\\
To explore Power-Law inflation, we assume $a=a_{0}t^{h_{0}}$, where
$h_{0}$ is a positive constant. Therefore, we have
\begin{equation}
H=\frac{h_{0}}{t},\,\,\dot{H}=-\frac{h_{0}}{t^{2}},\,\,\ddot{H}=
\frac{2h_{0}}{t^{3}},\,\,\dddot{H}=-\frac{6h_{0}}{t^{4}}.
\end{equation}
Note that the Gauss-Bonnet invariant in this case is as follows:
\begin{equation}
G=24\frac{h_{0}^{3}}{t^{4}}(h_{0}-1).
\end{equation}
Also we use the following approximate relations which work at the
inflationary epoch:
\begin{equation}
f(G)\approx \frac{1}{c_{1}}\big(1-\frac{c_{2}}{c_{1}}G^{-n}\big),
\end{equation}
\begin{equation}
f'(G)\approx \frac{n c_{2}}{c_{1}^{2}}G^{-(n+1)},
\end{equation}
\begin{equation}
f''(G)\approx \frac{-n(n+1) c_{2}}{c_{1}^{2}}G^{-(n+2)},
\end{equation}
\begin{equation}
f'''(G)\approx \frac{n(n+1)(n+2) c_{2}}{c_{1}^{2}}G^{-(n+3)}.
\end{equation}
 By substituting above approximate relations for $f(G)$ and its derivatives and (21) in (19),
 one can obtain
\begin{equation}
 h_{0}=\frac{2n+1}{2},
 \end{equation}

 \begin{equation}
a_{0}=\Bigg\{\big(\frac{2}{3}\big)^{(n+1)}\frac{2n}{(2n+1)^{3(n+1)}(2n-1)^{(n+2)}}\frac{c_{2}}{c_{1}^{2}}\frac{
\alpha\varepsilon^{2}+\beta\varepsilon b \tilde{g}^{2}+\gamma (b
\tilde{g}^{2})^{2}}{(4n^{2}-1)\varepsilon+2(2n+1)b
\tilde{g}^{2}}|B_{0}|^{2}\kappa^{2}\Bigg\}^{\frac{1}{4}},
 \end{equation}
 where
 $\alpha=32n^{5}+384n^{4}+\frac{2162}{3}n^{3}+\frac{1685}{3}n^{2}+\frac{607}{3}n+\frac{82}{3}$,\,\,
 $\beta=-\big(416n^{5}+1200n^{4}+\frac{4564}{3}n^{3}+\frac{3106}{3}n^{2}+\frac{1088}{3}n+\frac{149}{3}\big)$
 and
 $\gamma=160n^{5}+480n^{4}+\frac{2066}{3}n^{3}+\frac{1589}{3}n^{2}+\frac{607}{3}n+\frac{88}{3}$.\\
 We see that if $n\gg 1$ then $h_{0}$ becomes very large and so the power-law inflation can be
 realized. From this discussion, we conclude that non-minimally coupled YM field with $f(G)$
 gravity can be seen as a source of inflation in the early universe.
 This result is the same as that of non-minimal Maxwell- $f(G)$ gravity
 [31].\\
 We note that in this paper we considered only the case in which the values of the terms proportional
 to $f'(G)$, $f''(G)$ and $f'''(G)$ in equations (17) and (18) are dominant to the values of the term
 proportional to $\big(1+f(G)\big)$. Among the terms proportional to $f'(G)$, $f''(G)$ and $f'''(G)$, the
term proportional to $f'(G)$ is dominant and its value from equation
(24) is of order $f'(G) H^{4}\approx \frac{n c_{2}}{c_{1}^{2}}
H^{4}G^{-(n+1)}$. Also using (21) and (22) one can see that $G$ is of
order $20 H^{4}$. Now the condition for dominance $f'(G)$ in the
source term would be $\big(1+f(G)\big)/\Big[f'(G) H^{4}\Big]\sim
\frac{20}{n}\frac{c_{1}}{c_{2}}G^{n}\ll 1$. This leads to extremely
small $\big(\frac{c_{1}}{c_{2}}\big)$ because at the inflationary
epoch $G\gg 1$ and $n\gg 1$. In such a case, in equation (19) the
value of right hand side which is order
$\kappa^{2}f'(G)H^{4}\frac{|B_{0}^{a}|^{2}}{a^{4}}$ can be order
$H^{2}$. So, the right and left hand sides of equation (19) can
balance  each other and there is no contradiction between our result
and equation (19). In the opposite case i.e. if the term
proportional to
 $\big(1+f(G)\big)$ is dominant, the power-law inflation cannot be
 realized. Our reasoning is as follows: in this case the equations (17) and (18) can be written approximately
 as $H^{2}\approx \frac{\kappa^{2}}{6}\big(1+f(G)\big)\frac{|B_{0}^{a}|^{2}}{a^{4}}$ and
 $2 \dot{H}+3H^{2}\approx
 \frac{\kappa^{2}}{6}\big(1+f(G)\big)\frac{|B_{0}^{a}|^{2}}{a^{4}}$,
 respectively. Therefore one finds from equations (17) and (18) that
 $H^{2}$ and $2 \dot{H}+3H^{2}$ are of the same order and their
 difference $2 \dot{H}+2H^{2}$ must be much smaller than $H^{2}$.
 Then in equation (19) the quantity
$\Big(\dot{H}+\frac{\varepsilon}{\varepsilon-b\tilde{g}^{2}}H^{2}\Big)$
must balance with much smaller quantity than
 $\Big(\kappa^{2}\big(1+f(G)\big)\frac{|B_{0}^{a}|^{2}}{a^{4}}\Big)$,
 so
 $\Big(\dot{H}+\frac{\varepsilon}{\varepsilon-b\tilde{g}^{2}}H^{2}\Big)/H^{2}=
 \frac{\varepsilon}{\varepsilon-b\tilde{g}^{2}}-\frac{1}{h_{0}}\ll
 1$ and this leads to $h_{0}\ll 1$ because $\varepsilon ,b>0$. One can
 see that power-law inflation cannot be realized. From above discussion we see that power-law inflation can be
realized due to not the term proportional to $\big(1+f(G)\big)$ but
the term proportional to $f'(G)$ namely a non-minimal YM field
coupling.\\
 Finally, we note the following points. If one compares our result in
 equation (27) with those in non-minimal YM-$f(R)$ gravity in Ref.
 [24] can see that in our model when $n>\frac{3}{2}$ the universe is
 accelerating but in non-minimal YM-$f(R)$ gravity the expansion of
 the universe is accelerating for $n>3$. In addition, $f(G)$
 gravity is inspired from a fundamental theory such as string
 theory, so the non-minimal coupling of YM field with $f(G)$ may
 induce more interest than YM-$f(R)$ gravity.

\section{Inflation and late-time acceleration in modified Gauss-
Bonnet gravity framework} In this section, we consider a non-
minimally coupled YM field in the framework of modified Gauss-
Bonnet gravity
proposed in Ref. [32].\\
We describe the model by the following action:
\begin{eqnarray}
S=\int d^{4}x
\sqrt{-g}\Bigg[\frac{1}{2\kappa^{2}}\big(R+F(G)\big)-\frac{1}{4}\big
(F_{\mu\nu}^{a}F^{a\mu\nu}+f(G)F_{\mu\nu}^{a}F^{a\mu\nu}\big)\Big[1+b\tilde{g}^{2}\ln
\Big|\frac{-(1/2)F_{\mu\nu}^{a}F^{a\mu\nu}}{\mu^{4}}\Big|\Big]\Bigg],\nonumber\\
\end{eqnarray}
Note that in this case $F(G)$ is the modified part of gravity and it
is different from  $f(G)$ in the last term in the action (1). By
choosing FRW metric (8), the $(0,0)$ component and sum of $(i,i)$
components of equation of motion for $g_{\mu\nu}$, have the
following forms respectively:
\begin{equation}
H^{2}-\frac{1}{6}(GF'(G)-F(G))+4H^{3}\dot{G}F''(G)=\frac{\kappa^{2}}{3}T_{00}^{eff},
\end{equation}
and
\begin{equation}
2\dot{H}+3H^{2}+\frac{1}{2}(GF'(G)-F(G))-4H^{2}(\ddot{G}F''(G)+\dot{G}^{2}F'''(G))=-\kappa^{2}T_{ii}^{eff},
\end{equation}
where $T_{\mu\nu}^{eff}$ is given by equation (5). As the previous
section, we neglect the contribution of electric field and spatial
derivatives of $F_{\mu\nu}^{a}F^{a\mu\nu}$. Therefore, from
equations (30) and (31), one can obtain
$$\dot{H}+\frac{\varepsilon}{\varepsilon-b\tilde{g}^{2}}H^{2}+\frac{1}{6}(GF'(G)-F(G))\Big(\frac{2\varepsilon
-3b\tilde{g}^{2}}{\varepsilon-b\tilde{g}^{2}}\Big)-2H^{2}F''(G)\Big[\ddot{G}+H\dot{G}\Big(\frac{\varepsilon
-3b\tilde{g}^{2}}{\varepsilon-b\tilde{g}^{2}}\Big)\Big]\hspace{2cm}$$
$$-2H^{2}\dot{G}^{2}F'''(G)=\hspace{14cm}$$
$$\kappa^{2}\Bigg[4f'(G)\Big\{7\dot{H}H^{2}\varepsilon-\Big[
\Big(\frac{5\varepsilon-9b\tilde{g}^{2}}{\varepsilon-b\tilde{g}^{2}}\Big)\varepsilon-8b\tilde{g}^{2}\Big]H^{4}\Big\}+48f''(G)
\Big\{3\ddot{H}H^{5}(\varepsilon+b\tilde{g}^{2})-6H^{4}\dot{H}^{2}(\varepsilon-3b\tilde{g}^{2})$$
$$+4\dot{H}H^{6}(7
\varepsilon-b\tilde{g}^{2})-\big(8\ddot{H}\dot{H}H^{3}+6\dot{H}^{3}H^{2}
+\dddot{H}H^{4}\big)(\varepsilon-b\tilde{g}^{2})\Big\}+24f'''(G)
 \big(\ddot{H}H^{2}+2H\dot{H}^{2}\hspace{1.5cm}$$
\begin{equation}
+4\dot{H}H^{3}\big)^{2}(\varepsilon-b\tilde{g}^{2})\Bigg]\frac{|B_{0}^{a}|^{2}}{a^{4}}.\hspace{12cm}
\end{equation}
Here we take $F(G)$ from Ref. [32],
\begin{equation}
F(G)=\frac{(G-G_{0})^{2n+1}+G_{0}^{2n+1}}{c_{3}+c_{4}\big((G-G_{0})^{2n+1}+G_{0}^{2n+1}\big)},
\end{equation}
where $c_{3}$, $c_{4}$ are constants and $n$ is a positive integer.
$G_{0}$ correspond to the present value of the Gauss-Bonnet
invariant. The typical property of such theory is the presence of
effective cosmological constant epochs in such away that early-time
inflation and late-time cosmic acceleration are naturally unified
within single model.\\
Since $F'(G)=0$ when $G=G_{0}$ and $G=\infty$, $F(G)$ can be
regarded as an effective cosmological constant. We may consider
$F(\infty)$ as the cosmological constant for the inflationary stage
and $F(G_{0})$ as that at the present time,
\begin{equation}
\lim_{G\rightarrow\infty}F(G)=\frac{1}{c_{4}}=\Lambda,
\end{equation}
\begin{equation}
F(G_{0})=\frac{G_{0}^{2n+1}}{c_{3}+c_{4}G_{0}^{2n+1}}=2G_{0}.
\end{equation}
From the above equations, we find
\begin{equation}
c_{3}=\frac{G_{0}^{2n}}{2}-\frac{G_{0}^{2n+1}}{\Lambda}\approx\frac{G_{0}^{2n}}{2},\,\,\,\,\,\,\,\,\,\,\,
c_{4}=\frac{1}{\Lambda},
\end{equation}
because $\frac{G_{0}}{\Lambda}\ll1$.\\
Also, $f(G)$ is given by
\begin{equation}
f(G)=-\frac{(G-G_{0})^{2m+1}+G_{0}^{2m+1}}{c_{5}+c_{6}\big((G-G_{0})^{2m+1}+G_{0}^{2m+1}\big)},
\end{equation}
where $c_{5}$, $c_{6}$ are constants and $m$ is a positive integer.
At the inflationary epoch we can use the following approximate
relations:
\begin{equation}
F(G)\approx\frac{1}{c_{4}}\Big[1-\frac{c_{3}}{c_{4}}(G)^{-(2n+1)}\Big],
\end{equation}
and
\begin{equation}
f(G)\approx-\frac{1}{c_{6}}\Big[1-\frac{c_{5}}{c_{6}}(G)^{-(2m+1)}\Big].
\end{equation}
Because $G\rightarrow\infty$ at the inflationary stage and also
$\lim_{G\rightarrow\infty}F(G)=\Lambda$ and
$\lim_{G\rightarrow\infty}f(G)=const$, equation (32) at this epoch is
reduced to
\begin{equation}
\dot{H}+\frac{\varepsilon}{\varepsilon-b\tilde{g}^{2}}H^{2}=\frac{\Lambda}{6}\Big(\frac{2\varepsilon
-3b\tilde{g}^{2}}{\varepsilon-b\tilde{g}^{2}}\Big).
\end{equation}
It follows from above equation that
\begin{equation}
a(t)\propto exp(\frac{\Lambda}{3})^{\frac{1}{2}}t,
\end{equation}
Hence exponential inflation can be realized. Thus, we conclude that
the terms in $F(G)$ on the left hand side of Eq. (32) can be a
source of inflation, in addition to $f(G)$ on the right hand side of
Eq. (32). Note that if we do not consider the contribution of the
terms in $F(G)$ to inflation, equation (32) is reduced to equation
(19). In this case, substituting $a=a_{0}t^{h_{0}'}$ and the approximate
expressions of $f'(G)$, $f''(G)$ and $f'''(G)$ derived from equation
(39) into equation (32) leads to $h_{0}'=\frac{4m+3}{2}$. Hence if
$m\gg 1$, $h_{0}'$ becomes much
larger than unity and power-law inflation can be realized.\\
We emphasize that there are two sources of inflation in the present
model, one from the modified part of gravity $F(G)$ and the other
from the non-minimal coupling of YM field with $f(G)$. Indeed, in
this model even if the value of $\Lambda$ is so small that the
modification of gravity cannot contribute to inflation then
inflation can be realized due to the non-minimal gravitational
coupling of the YM field. This is an important cosmological
consequence of the present model.\\
At the present time, because $G-G_{0}\ll1$, if $m>n$, $f(G)$ becomes
constant more rapidly than $F(G)$ in the limit $G\rightarrow G_{0}$.
For such a case, when $G\rightarrow G_{0}$ Eq. (32) leads to
\begin{equation}
\dot{H}+\frac{\varepsilon}{\varepsilon-b\tilde{g}^{2}}H^{2}=\frac{G_{0}}{3}\Big(\frac{2\varepsilon
-3b\tilde{g}^{2}}{\varepsilon-b\tilde{g}^{2}}\Big),
\end{equation}
so, from this equation one can obtain
\begin{equation}
a(t)\propto exp(\frac{2G_{0}}{3})^{\frac{1}{2}}t,
\end{equation}
so that the late-time acceleration of the universe can be realized.
These results are also in agreement with the results of Ref. [31]
where non-minimal Maxwell- $F(G)$ gravity has been investigated. We
mention that even if the value $G_{0}$ is so small that the
modification of gravity cannot contribute to the late-time
acceleration of the universe, the late-time accelerated expansion
can be realized due to the non-minimal coupling of the YM field.\\
Indeed our results are the generalization of the results for
non-minimal Maxwell theory with the coupling of the electromagnetic
field to a function of Gauss-Bonnet invariant [31].In here we
considered a non-Abelian gauge field (the YM field) non-minimally
coupled with $f(G)$ gravity. The YM fields are indispensable to
particle physics, there is no room for adjusting the functional form
of the Lagrangian as it is predicted by quantum field theory. As a
model for the cosmic dark energy, it has no free parameters except
the present cosmic energy scale and the cosmic evolution only
depends on the initial conditions [33].

\section{Conclusion}
To summarize, the non-minimal gravitational coupling of YM field
with Gauss-Bonnet invariant function, $f(G)$, has been considered in
Friedmann-Robertson-Walker background metric. Such a non-minimal
coupling has been examined in the framework of general relativity.
We have shown that power law inflation can be realized due to
non-minimal coupling of YM field in this model which is described by
action (1). We have also studied cosmology in non-minimally coupled
YM field in the framework of modified Gauss-Bonnet gravity, $F(G)$.
It has been shown that both inflation and late-time acceleration of
the universe can be realized in such a model proposed in Ref.
[32].\\
Clearly, more checks of this theory such as stability/instability of
inflation should be done in order to conclude if the model is
realistic or not. The conditions for stability of $f(G)$ rarity
have been derived in [17]. It has been shown that the condition
$\frac{d^{2}f}{dG^{2}}>0$ needs to be fulfilled in order to ensure
the stability of a late-time de-sitter solution as well as the
existence of standard radiation and matter dominated epochs [17].
Studying stability/instability conditions for our model and models
of this kind such as the Maxwell-$f(G)$ model [31] will be our plan for future works.\\
It is also interesting to extend our formulation for more
complicated theories. For instance one can investigate non-minimal
coupling of YM Lagrangian with non-local $f(G)$ gravity proposed in
[39] or one can study our model in the case that instead of $f(G)$
gravity an action with higher order string loop corrections
replaced. This kind of superstring inspired action has been
considered in [40]. \\

\textbf{Acknowledgements} \\
The authors are indebted to the anonymous referees for their
comments that improved the paper drastically.

\end{document}